\documentclass[12pt]{article}
\usepackage{graphics}
\begin{document}
\begin{center}
{\bf Elastic constants of nematic liquid crystals of uniaxial symmetry}
\end{center}
\begin{center}
{\ Amit Srivastava and Shri Singh} \\
{\it Department of Physics, Banaras Hindu University, Varanasi-221005, India}
\end{center}
\begin {center}
{\bf Abstract}
\end{center}
We study in detail the influence of molecular interactions on the
 Frank elastic constants of uniaxial nematic liquid crystals composed 
of molecules of cylindrical symmetry. A brief summary of the status of theoretical 
development for the elastic constants of nematics is presented. Considering 
a pair potential having both repulsive and attractive parts numerical calculations
 are reported for three systems MBBA, PAA and 8OCB. For these systems the length-to-width ratio ${x_0}$
 is estimated from the experimentally proposed structure of the molecules. 
The repulsive interaction is represented by a repulsion between hard ellipsoids of revolution
(HER) and the attractive potential is represented by the quadrupole and dispersion interactions.
From the numerical results we observe that in the density range of nematics the contribution
of the quadrupole and dispersion interactions are small as compared to the repulsive HER interaction.
The inclusion of attractive interaction reduces the values of elastic constants
ratios. The temperature variation of elastic constants ratios are reported and compared 
with the experimental values. A reasonably good agreement between theory
and experiment is observed.

\section{\bf Introduction}
In a previous paper \cite{1} (here after reffered to as I), we developed a theory based on the density
functional formalism \cite{2} for the deformation free-energy of any systems with continuous broken
symmetry, and applied the theory to derive expressions for the elastic constants of nematic 
and smectic-A phases of uniaxial ($D_{\infty h}$) symmetry. These expressions of elastic constants are written in 
terms of order parameters that characterize the nature and amount of ordering in the phase and 
the structural parameters involving the generalized spherical harmonic coefficients of the direct pair correlation 
function of an effective isotropic liquid, the density of which is determined using a criterion of the weighted density
functional formalism \cite{3}.

In the present paper we restrict ourselves to the uniaxial nematic phase. According to the 
continuum theory \cite{4}, the bulk elastic properties of nematics can be described by three elastic
constants, associated with the restoring forces opposing splay ($K_1$), twist ($K_2$) and 
bend ($K_3$) distortions. The distortion free energy density is written as
\begin{equation}
\frac{1}{V}\Delta{A_e} = \frac{1}{2}[K_1({\bf{\nabla}}\cdot{\bf{\hat {n}}})^2+K_2({\bf{\hat{n}}}\cdot{\bf{\nabla}}\times{\bf{\hat{n}}})^2
+K_3({\bf{\hat{n}}}\times{\bf{\nabla}}\times{\bf{\hat{n}}})^2]
\end{equation}
where ${\bf{\hat n}}$, the director, indicates the preferred direction of the long-axes of the molecules.
These elastic constants play an important role in characterizing liquid crystal displays. It is difficult to measure experimentally
the absolute values of these elastic constants \cite{5,6,7,8}. Their ratios ${K_2}/{K_1}$and 
${K_3}/{K_1}$ can be measured more accurately \cite{8,9,10}. For high duty liquid crystal
displays \cite{11}, the ratio ${K_3}/{K_1}$ is desired to be as small as possible. The elastic moduli are temperature and 
density dependent. The dependence on the density is pronounced. A number of measurements are reported which
show that $K_1$ and $K_2$ have weak temperature dependences whereas $K_3$ rapidly increases
with temperature and that when the nematic-smectic A transition temperature is approached from above, $K_1$ does not show any sharp
change but $K_2$ and $K_3$ increases anomalously.

The present paper is organized as follows: In sec.2, we describe, in brief, the status of the theoretical development
for the elastic constants of the nematics and summarize the working equations used in the calculation.
The numerical evaluation and results are presented in sec.3

\section{\bf Theoretical development and working equations}
First we shall comment, in brief, on the status of the theoretical development for the elastic constants of nematics.
A detail of these works are well documented elsewhere \cite{10,12}

In the Landau-de-Gennes theory the free-energy density is assumed to be an analytic function of the order parameter tensor. To the extent
that the order parameter is small, the free-energy density is expressed as an expansion in its various orders and gradient terms. The elastic 
free-energy density upto second order in order parameter is written in terms of two elastic constants
$L_1$ and $L_2$ which are related with the Frank elastic constants as

\begin{equation}
K_1 = \frac{q}{2}(L_1 + \frac{1}{2}L_2){\overline{P_2}}^2
\end{equation}
\begin{equation}
K_2 = \frac{q}{2}L_1\overline{P_2}^2
\end{equation}
and 
\begin{equation}
K_1 = K_3
\end{equation}
Thus the elastic constants moduli $K_i$ vary with temperature like $\overline{P_2}^2$. The prediction $K_1$=$K_3$
is not consistent with the experimental observation. However, this is an artifact of the derivation in which only the gradient
terms of $\overline{P_2}$ order parameter have been considered. If $\overline{P_4}$ term is also included in the free-energy expansion, all 
the three elastic constants will be different. Using an expansion of the intermolecular potential in spherical harmonics
Priest \cite{14} also arrived at the same conclusion. For the special case of dispersion forces, Nehring and Saupe \cite{13}
calculated (up to the second order $\overline{P_2}$) the ratio of the elastic constants
\begin{center}
${K_1} : {K_2} : {K_3}$ = $ 5:11:5$
\end{center}
Apart from the temperature dependence of the elastic constants, via $\overline{P_2}$ order parameter, there is also a
dependence on the density. Priest \cite{14} showed that the deviation from the equality $K_1$ = $K_3$ are related in a simple way to the ratio
${\overline{P_4}}/{\overline{P_2}}$ and

\begin{equation}
{K_1}/{\overline{K}} = \Delta - \Delta'({\overline{P_4}}/{\overline{P_2}}) + \dots 
\end{equation}
\begin{equation}
{K_2}/{\overline{K}} = -2\Delta - \Delta'({\overline{P_4}}/{\overline{P_2}}) + \dots
\end{equation}
\begin{equation}
{K_3}/{\overline{K}} = \Delta + 4\Delta'({\overline{P_4}}/{\overline{P_2}}) + \dots
\end {equation}

where $\overline{K}$= $\frac{1}{3}$ ({${K_1}$+ ${K_2}$+ ${K_3}$}). For the case of hard spherocylinders, the constants {$\Delta$}
and {$\Delta'$} were found to depend on the length-width ratio of the molecules. Most of the hard rod models \cite{10} are strictly
valid only for the very long and thin rods and they usually predict too large a value for $K_3$ and cannot reproduce the temperature
dependences of the elastic constants.

Several workers \cite{10} have evaluated the elastic constants for the van der Walls type potential described by hard
 spherocylinders with superimposed attractive interactions.While Stecki and Poniewierski \cite{15} treatment is based on the the direct
correlation function  approach, mean field (MF) approximation has been adopted by Kimura et. al. \cite{16}. These 
authors found that {$K_3> K_1> K_2$} and that the temperature dependence of their ratios is in accordance with
the experiment. Using generalized van der Walls (GVDW) theory \cite{17}, which couples the contributions of the short-and long-range 
pair potentials, the elastic moduli were evaluated in a model based on the distributed harmonic forces \cite{18} between the molecules. 
In this work both the repulsive and attractive forces have been considered as distributed 
along the molecules. This approach disregards the temperature dependence of elastic constants 
and assumes perfect orientational order i.e. 
$\overline{P_2}$ =$\overline{P_4}$ =1. It is a variant on the ideas of Gelbart and co-workers \cite{17} who studied the combined
effect of repulsive and attractive forces. Zakhrov \cite{19} evaluated the elastic constants and order parameters by using a 
theory that is  based on the method of conditional distribution (MCD) \cite{20}. This method introduces a concept of reduced distribution 
functions which obey infinite chains of integro-differential equations. For an arbitrary equations of the chains based on the concept of the 
mean-force potential (MFP) \cite{21,22} a truncated procedure was adopted. The numerical solution exhibits certain 
qualitative features: The order parameters decrease with increasing volume and temperature. While the observed value of 
${K_1}/{\overline{K}}$ and ${K_2}/{\overline{K}}$ increases with increasing volume, the values of ${K_3}/{\overline{K}}$
decreases with it. ${K_1}/{\overline{K}}$ increases strongly with the length-width ratio and ${K_2}/{\overline{K}}$
decreases with it, and $0.5 <{K_3}/{K_1}<3.0$ and $0.5<{K_2}/{K_1}<0.8$ 

Several workers have considered the application of density functional theory to study the elastic properties of nematics. A detail account of these works are summarized elsewhere by one of us
\cite{10}. In this theory exact expression for the elastic free- energy is obtained in terms of order parameters and direct correlation functions. The correlation of the ordered
phases are, in general, not known and hence to be approximated. The functional Taylor expansion is performed to obtain 
the n-particle direct correlation function of an inhomogeneous system from the n- and higher-order direct correlation functions of an uniform system. 
The elastic constants of nematics were calculated for a number of model systems using approximate forms of the pair 
correlation function of the medium. However, none of these approximate forms of the pair correlation function gives the structure
of the medium correctly, and so the result reported are not expected to be accurate.

A unified molecular theory was developed by Singh et. al. \cite{1} for  the deformation free-energy of ordered molecular phases. This theory is based on
the weighted density functional formalism \cite{3} and writes the exact expressions for the elastic constants, in terms of integrals involving spherical harmonic coefficients of the direct pair correlation function 
of an effective isotropic liquid . Adopting the procedure as outlined in I, we express the elastic 
free- energy density for a uniaxial nematic phase of axially symmetric molecules. 
\begin{eqnarray}
\frac{1}{V}\beta\Delta{A_e}[\rho] & = & -\frac{1}{2}{\rho_{n}}^2{\sum_{l_1l_2l}}'\sum_{m}
[\frac{(2l_{1}+1)(2l_{2}+1)}{(4\pi)^{2}}]^{\frac{1}{2}}
{\overline{P}}_{l_1}{\overline{P}}_{l_2} C_g(l_1 l_2 l, o m n) \nonumber \\  
& &  \int{d}{\bf {r}_{12}}C_{l_1l_2l}({\bf {r}_{12}})[(\frac{4\pi}{2l_{2}+1})
^{\frac{1}{2}}Y_{l_2m}(\Delta\chi({\bf {r}_{12}}))- 1]
{Y^{*}_{lm}}({\bf{\hat{r}}}_{12})
\end{eqnarray}
where $\rho_n$ is the nematic number density and $\overline{P_l}$ are the Legendre polynomial order parameters. The prime on the 
summation indicates that $l_1$ and $l_2$ are even ${\bf {\hat{r}}}_{12}$ =  ${\bf {r}}_{12}/| {\bf {r}}_{12}|$
is unit vector along the intermolecular axis, $C_g$ are the Clebsch-Gordan coefficients, ${C_{l_1l_2l}}$ are the spherical harmonic
coefficients of the direct pair correlation function of an isotropic liquid, and $\Delta{\chi}({\bf {r}}_{12})$represents the angle between the principal director
at $\bf {r_1}$ and $\bf {r_2}$. Confining the variation of director at $ \bf {r_2}$ in a plane, the $Y_{l_2m}(\Delta\chi({{\bf {r}}_{12}}))$  is 
expressed in terms of the distortion angle which is assumed to be small. Performing the angular integration over
 $\bf {\hat{r}}_{12}$ and comparing eq. (8) with eq. (1), the following expressions for the elastic constants of 
uniaxial nematic phase composed of cylindrically symmetric molecules are obtained \cite{1},  

\begin{equation}
\beta {K_i}= {\sum_{l_1,l_2}}'\beta {K_i}(l_1, l_2)
\end{equation}

The explicit expressions for the first few terms of the series can be written as

\begin{equation}
\beta K_1(2,2) = (\frac{5}{4\pi})^\frac{1}{2}\rho_{n}^{2}{\overline{P}}_{2}^{2}[\frac{1}{2}J_{220} - \frac{1}{\sqrt{14}}J_{222}]
\end{equation}
\begin{equation}
\beta K_1(2,4) = - \frac{3}{4}(\frac{5}{\sqrt{14\pi}})\rho_n^{2}{\overline{P}}_{2}{\overline{P}}_{4}J_{242}
\end{equation}
\begin{equation}
\beta K_1(4,4) = (\frac{5}{4\pi})^\frac{1}{2}\rho_{n}^{2}{\overline{P}}_{4}^{2}[\sqrt{5}J_{440} - \frac{13}{2\sqrt{77}}J_{442}]
\end{equation}
\begin{equation}
\beta K_2(2,2) = (\frac{5}{4\pi})^\frac{1}{2}\rho_{n}^{2}{\overline{P}}_{2}^{2}[\frac{1}{2}J_{220} + \sqrt{\frac{2}{7}}J_{222}]
\end{equation}
\begin{equation}
\beta K_{2}(2,4) = \beta K_{1}(2,4)
\end{equation}
\begin{equation}
\beta K_2(4,4) = (\frac{5}{4\pi})^\frac{1}{2}\rho_{n}^{2}{\overline{P}}_{4}^{2}[\sqrt{5}J_{440} + \frac{47}{2\sqrt{77}}J_{442}]
\end{equation}
\begin{equation}
\beta K_3(2,2) = \beta K_1 (2,2)
\end{equation}
\begin{equation}
\beta K_3(2,4) =-\frac{4}{3}\beta K_1(2,4) 
\end{equation}
and
\begin{equation}
\beta K_3(4,4) = (\frac{5}{4\pi})^\frac{1}{2}\rho_{n}^{2}{\overline{P}}_{4}^{2}[\sqrt{5}J_{440} + \frac{17}{\sqrt{77}}J_{442}]
\end{equation}

In this expressions the structural parameter $J_{l_1l_2l}$ are defined as
\begin{equation}
J_{l_1l_2l} = \int r_{12}^4 dr_{12}C_{l_1l_2l}(r_{12}) 
\end{equation}

\section{\bf Calculation and Results}
As obvious from eq. (9) the evaluation of elastic constants requires the values of density, order parameters $\overline{P_2}$
and $\overline{P_4}$ (as a function of temperature), the structural parameters ${J_{l_1l_2l}}$ ( as a function of density and
 length-width ratio $x_0$) and potential parameters. In I it was shown that the dipole-dipole and dipole-quadrupole
interactions do not contribute to the free energy (and hence elastic constants) of the uniaxial mesophases. For the numerical calculation we consider a system of molecules with prolate ellipsoidal
symmetry interacting via a pair potential,

\begin{equation}
u({\bf r_{12}}, \Omega_{1}, \Omega _{2}) = (u_{HER}+u_{qq}+u_{dis})({\bf r_{12}}, \Omega_1, \Omega_2)
\end{equation}

where $u_{HER}$ represents the repulsion between hard ellipsoids of revolution (HER) and the subscripts qq and dis indicate
respectively, the interactions arising due to the quadrupole-quadrupole and dispersion forces. The explicit from of these 
interactions are given in I.

For evaluating structural parameters , $J_{l_1l_2l}$, as a function of density and $x_0$, the values of spherical harmonic coefficients 
$C_{l_1l_2l}$ for the potential model (20) are needed. The $C_{l_1l_2l}$ can be obtained by solving the Ornstein- Zernike
(OZ) equation, using the Percus-Yevick (PY) closure relations. As this evaluation is difficult, in a calculation only a finite 
number of spherical harmonic coefficients for any orientation dependent function can be handled.
It has been shown that for the potential model (20) the inclusion of all the 
harmonics up to indices $l_1$, $l_2$= 4 makes the series fully convergent.

In this paper we study in detail the influence of HER, qq, dispersion interactions on the elastic 
properties of three uniaxial nematic liquid crystals: p-azoxyanisole (PAA), 
N-p-methoxybenzylidene-p-butylaniline (MBBA) and 
$4'$-pn-octyloxy-4-cyanobiphenyl $(8OCB)$.
For these systems we have estimated the values of length-width ratio $x_0$ from the proposed structure of molecules
as determined by $^{14}N$  nuclear quadrupole resonance \cite{23} and line shape \cite{24} studies. The estimated values of $x_0$ are 2.68, 2.87 and 4.88 for MBBA
, PAA and 8OCB, respectively. The PY closure relation have been solved by Ram and Singh \cite{25} for the $g_{HER}$, $h_{HER}$
and $C_{HER}$ harmonics for $x_0$=3.00, 3.25, 3.50 and 4.00. Taking their results we estimated in I C-harmonics for the 
quadrupolor and dispersion interactions. Adopting similar procedure we have evaluated here these harmonics for 
$x_0$ = 2.68, 2.87, and 4.88. With known C- harmonics we evaluated the values of structural parameters as a function 
of reduced density $\rho^{*}$( =$\rho_{n}{d_{0}^{3}})$ where $d_{0}$ is the molecular diameter.
In addition to $J_{l_1l_2l}$ we need the values of order parameters $\overline{P_2}$ and $\overline{P_4}$ as a function of temperature
and of potential parameters ${\epsilon_{0}}/{k}$, $d_0$ and quadrupole moment $\Theta$. Here $\epsilon_{0}$ is constants with unit of energy. In the calculation we use the values of 
$\overline{P_2}$ as determined by the experiments and estimate $\overline{P_4}$ as 
${\overline{P_4}}/{\overline{P_2}}$ $\cong{\overline{P_2}^{2}}$

We calculate the contribution of the individual terms of the series

\begin{equation}
K_i = K_i(2,2) + 2K_i(2,4) + K_i(4,4)
\end{equation}
for the HER, qq and dispersion interactions for three values of $x_{0}$( = 2.68(MBBA), 2.87(PAA) and 4.88 (8OCB)). A number of observations regarding relative contributions
of these interactions terms have been made from this calculation: In the density range of nematics the contribution 
of quadrupole and dispersion interactions are small as compared to the repulsive HER interaction. The absolute values
of elastic constants are sensitive to the values of molecular parameters. The inclusion of dispersion and quadrupole interactions reduce the values of elastic constants
ratios. For a given $x_0$, $K_3^{HER}(2,4)$ is positive where as $K_2^{HER}(2,4)$ and $K_1^{HER}(2,4)$ are negative. Consequently we find that 
$K_3^{HER} > K_1^{HER} > K_2^{HER}$ and the ratio (${K_3^{HER}}/{K_1^{HER}}) > ({K_2^{HER}}/{K_1^{HER}}$).
This result is in accordance with the simulation work \cite{27}. The absolute values of $K_i^{HER}$ increase linearly with the temperature and the values of their ratios do not 
change with temperature. In case of the quadrupole interaction for a
 given $x_0$ and quadrupole moment $\Theta$ the contribution of $K_i^{qq}(4,4)$is much smaller as compared to $K_i^{qq}(2,2)$. As 
the values of $\Theta$ increases the contribution of each individual terms of the series (21) and $K_i^{qq}$ increase significantly. The numerical values of $K_1^{qq}$ and $K_3^{qq}$ are positive where as $K_2^{qq}$ is negative. The influence of the dispersion interaction is small
as compared to the quadrupole interaction

The temperature dependence of the elastic constants are mainly due to the variation of $\overline{P_2}$ and $\overline{P_4}$
with temperature. We calculate the values of $K_i$ using the experimental values of $\overline{P_2}$. Using available experimental data of  $\overline{P_2}$
for PAA \cite{27} and MBBA \cite{28} we draw a smooth curve and the values of $\overline{P_2}$ used in the calculation correspond to this smooth curve. For 8OCB
we use the experimental values of $\overline{P_2}$ as measured by Madhusudana and Pratibha \cite{9}. We have found that the values of $\overline{P_4}$
for PAA as estimated from the relation $\overline{P_4}\cong{\overline{P_2}^3}$ are in good agreement with experimental data
\cite{29} of deuteriated PAA obtained from the coherent neutron scattering experiment. So in the calculation we use this estimated values of $\overline{P_4}$

Experimentally it is difficult to obtain the absolute values of elastic constants. It is the ratios ${K_2}/{K_1}$ and 
${K_3}/{K_1}$ which are usually measured more accurately. In addition to these ratios the other quantity which one finds 
accurately from the experiment is ${K_i}/{\overline{K}}$. In accordance with the experiment we observe that the ratio 
${K_3}/{K_1}$ decreases significantly with increasing temperature whereas the ratio ${K_2}/{K_1}$ is more or less independent
of the temperature. So in the following figures we show a comparison between the theoretical and experimental values of the ratios
${K_3}/{K_1}$, ${K_1}/{\overline{K}}$ and ${K_3}/{\overline{K}}$ for the PAA, MBBA and 8OCB. 

A comparison between the experimental \cite{30,31} and theoretical values of ratios ${K_3}/{K_1}$, ${K_1}/{\overline{K}}$ and ${K_3}/{\overline{K}}$
for PAA is shown in fig.1. It can be seen that the theoretical values are consistent with the experimental data. We have found that the inclusion of
quadrupole and dispersion interactions decreases the values of elastic constants ratios. Figure 1 shows that the ratio ${K_3}/{K_1}$
decreases significantly with temperature. A weak temperature dependence is found for the ratios ${K_i}/{\overline{K}}$.
As temperature increases the value of ${K_3}/{\overline{K}}$ decrease whereas ${K_1}/{\overline{K}}$ increases. As obvious from fig. 2 a similar trend in the variation of these ratios 
has been found for MBBA. In case of 8OCB fig.3 a similar but more pronounced variation of these ratios with temperature is observed. Further, as physically
expected near the nematic-smectic A transition a pronounced increase in the value of $K_3$ is clearly observed.

Table 1 shows the variation of elastic constants ratios with length-width ratio near nematic-isotropic transition temperature. It can
be seen that with $x_0$ the ratio ${K_3}/{K_1}$ and ${K_3}/{\overline{K}}$ increase whereas ${K_1}/{\overline{K}}$
and ${K_2}/{\overline{K}}$ decrease.

As mentioned in the text, the values of elastic constants are sensitive to the values of the structural parameters, order parameters and potential parameters. The available information about these 
parameters are not acurate. So as our knowledge of these parameters improves, more accurate values of elastis constants will result.

\begin{table}
{\small Table 1: Ratio of elastic constants of nematic liquid crystals near nematic-isotropic   \\  
                           transition temperature $T_{NI}$}\\
\begin{center}
\begin{tabular}{|c|c|c|c|c|} \hline 
$x_0$ & $\frac{K_3}{K_1}$ & $\frac{K_3}{\overline K}$ & $\frac{K_1}{\overline K}$ & 
$\frac{K_2}{\overline K}$  \\
\hline
2.68 & 1.211 & 1.276 & 1.053 & o.747 \\
\hline
2.87 & 1.458 & 1.34 & 0.919 & 0.739  \\
\hline
4.88 & 1.542 & 1.409 & 0.917 & 0.682  \\
\hline
\end{tabular}
\end{center}
\end{table}
 
\begin{figure}
\includegraphics{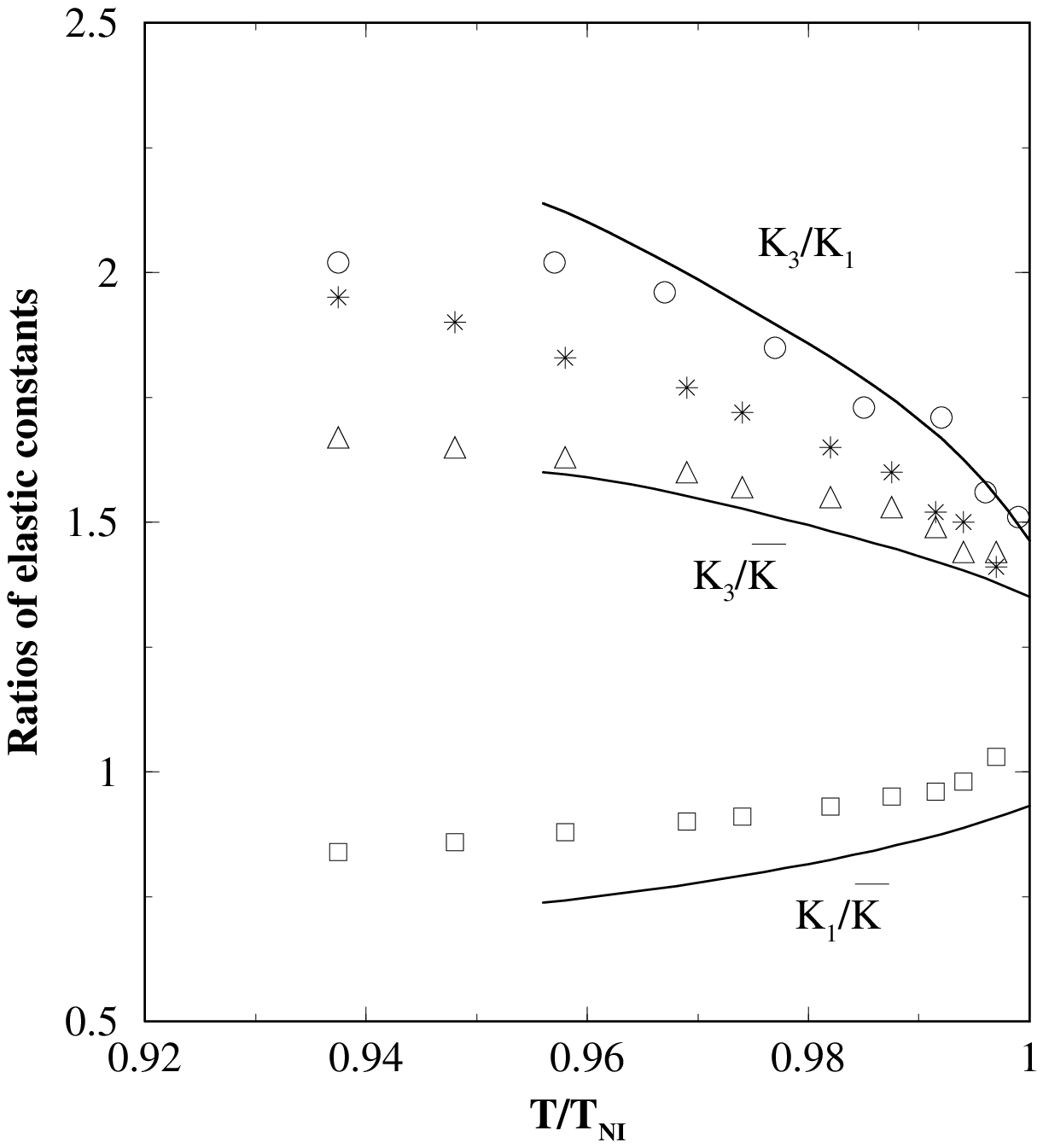}
\caption{Comparison between the calculated (solid lines) and experimental values
of the elastic constants ratios for PAA as a function of temperature. The experimental
ratios o[30], *[31] are shown as $K_3/K_1$ ; o and * ; $K_3/{\overline{K}}$ ; $\Delta$ and $K_3/{\overline{K}}$ ; square}
\end{figure}
\begin{figure}
\includegraphics{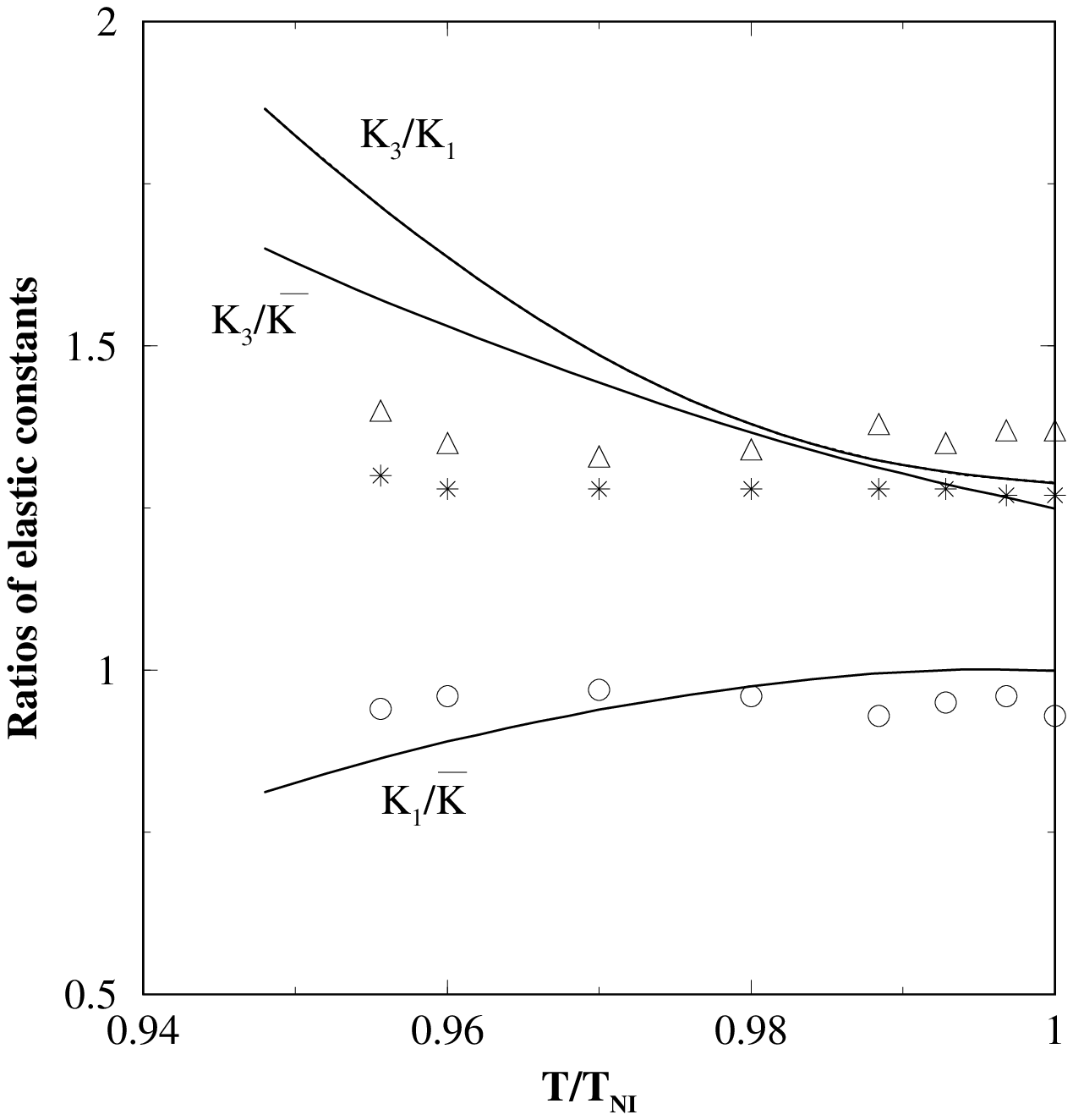}
\caption{Comparison between the calculated (solid lines) and experimental values [32
] of the elastic
constants ratios for MBBA as a function of temperature. The experimental ratios are shown as
$K_3/K_1$ : $\Delta$ ; $K_3/{\overline{K}}$ : *  and $K_1/{\overline{K}}$ : o } 
\end{figure}
\begin{figure}
\includegraphics{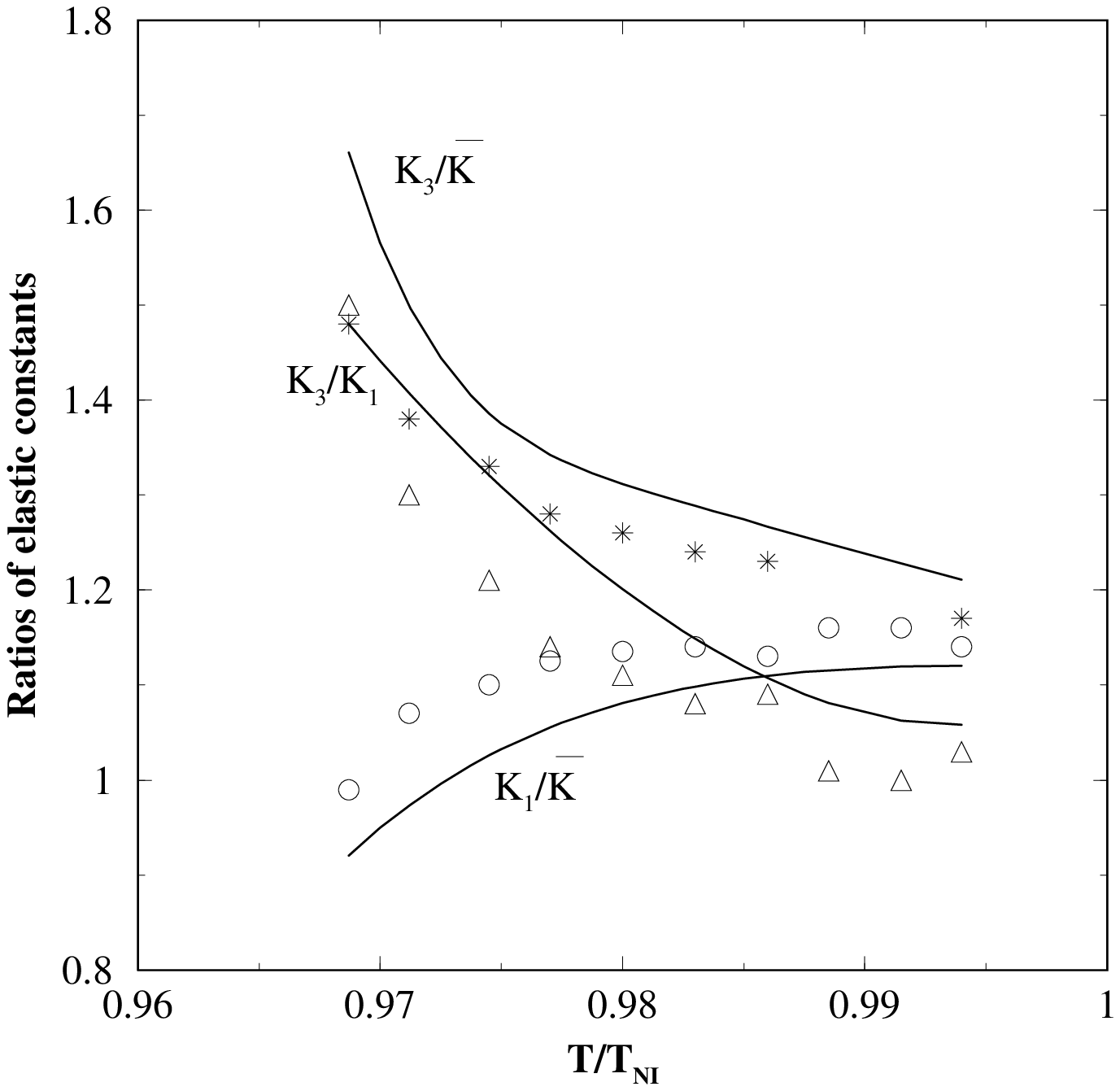}
\caption{Comparison between the calculated (Solid lines) and experimental[9]
values of elastic constants ratios for 8OCB as a function of temperature.
The experimental ratios are shown as $K_3/K_1$ : $\Delta$ ; $K_3/{\overline{K}}$: * and $K_1/{\overline{K}}$ : o }
\end{figure}

\end{document}